\documentclass[preprint,showpacs,amsmath,amssymb,aps]{revtex4}

\begin{document}
\title{Symmetry and inert states of spin Bose Condensates} 
\author{S.-K. Yip}
\affiliation{Insitute of Physics, Academia Sinica,
Nankang 115, Taipei, Taiwan}
\date{\today}
\begin{abstract}

We construct the list of all possible inert states
of spin Bose condensates for $S \le 4$.  In doing
so, we also obtain their symmetry properties.
These results are applied to classify line defects
of these spin condensates in zero magnetic field.

\end{abstract}
\maketitle

\section{Introduction}

Optical trapping of atomic Bose gases 
\cite{MIT,MIT2,Chang,Sengstock04,Griesmaier05}
has provided us with exciting new physical systems,
which have attracted significant theoretical interests
(e.g.
 \cite{Ho98,Machida,Ciobanu00,Martikainen,Ueda02,Isoshima06,Diener06,Santos06}).
 In constrast to
$^4$He, these Bosonic atoms possess finite hyperfine spins
(hereafter simply as spins, denoted by $S$),
 hence internal degrees of freedom.
Therefore in the Bose condensed phase, in contrast to $^4$He
where only the gauge symmetry is broken, the state can also have
broken spin-rotational symmetries 
(we ignore the case of fragmented states \cite{HoYip00} here).
The present paper is devoted to symmetry
considerations of these spin-condensates.
Knowing the broken symmetries, as well as the residual symmetries,
is essential in understanding the properties of these spin superfluids,
in particular topological excitations such as vortices.

The states that are expected to be realized are determined
by minimization of (free) energy of the system.
Such a program has been carried out by numerous authors
\cite{Ho98,Machida,Ciobanu00,Diener06,Santos06},
assuming that mean-field theory applies.
The minimization of this free energy is rather simple \cite{Ho98,Machida}
for $S=1$, already not quite trivial
\cite{Ciobanu00} (see also \cite{Mermin74}) for $S=2$,
and indeed rather complicated \cite{Diener06,Santos06} for 
$S=3$.  There are also a few attempts to find out
the residual symmetries of these minimum energy states found
\cite{Makela03,Barnett06}.  

On the other hand, symmetry considerations can actually help
in finding a special set of possible states known as "inert" states
for the systems.  These states are those where the wavefunction
characterizing the state is actually independent of the
(interaction) parameters that enter in the free energy that
is being minimized.  The form of these wavefunctions
follow from symmetry considerations alone, depending only
on the symmetries of the uncondensed (normal) state
(thus independent of whether mean-field theory has been assumed).
>From such considerations, the symmetries of these inert states
are thus also directly determined.  It is however, true that
this procedure only determines the possible stationary points
(relative minima, extrema or saddle points) of the free energy functional.
Additional calculations (in particular the actual form
of the free energy and the parameters which enter it)
 are needed to determine which state is
actually a stable minima.

Such a program has actually been carried out before in other
systems.  In superfluid $^3$He, Cooper pairing
occurs in angular momentum $L=1$ and spin $S=1$ channel.
Spatial and/or spin rotational symmetries are in general
broken in the superfluid states, and 
symmetry considerations \cite{Bruder86,VW} (and references therein)
have been used to find the inert states.
These considerations have also been applied to "unconventional" 
superconductivity in solid state systems, 
where the superconducting state has symmetries
lower than the underlying crystals, in particular
in the context of heavy fermions, e.g., \cite{VG,Ozaki}.
In this case, the normal state possesses only discrete rather
than continuous rotational symmetries.

In this paper, we shall carry out such a program for the spin
condensates, finding the inert states and their
symmetries.  We shall limit ourselves to the case of no
external magnetic field (linear and/or
quadratic Zeeman fields).  
%Thus the normal state is
%assumed to have complete three-dimensional spin-rotational
%SO(3), gauge U(1), and time-reversal
%symmetry $\Theta$.  
We discuss the general theory behind this work in Sec \ref{basics}.
Results up to $S=4$ are then presented in Sec \ref{results}.
We then apply these results to discuss vortices on
Sec \ref{vortices}.
Sec \ref{conclusions} provides a brief summary,
and some details are relegated to the appendices.

\section{\label{basics} Basic theory}

We here outline the basic theory behind this work.
The present procedure of finding the inert states
is based on a theorem by Michel \cite{Michel}.  This
theorem has been discussed and explained in, e.g.,
\cite{VW}.  Rather than restating this theorem 
in its full generality in abstract terms, we shall simply
describe it physically for the case of $S=1$ here.
Since $S=1$ has the same symmetry as angular momentum $L=1$,
we shall use these two terminologies interchangably.
In the uncondensed phase (the normal phase), the system
is invariant under all spatial rotation 
SO(3) and gauge transformation U(1), and time reversal $\Theta$.
(There is no need to consider inversion since we shall
always consider fixed $L$, so all states have definite inversion
symmetry).  We are interested in finding the stationary points
of a free energy $F[\Psi]$ which is a functional of $\Psi$.
$F[\Psi]$ is invariant under any transformation
$\Psi \to g \Psi$ where $g$ is any operation in the
group G $=$ SO(3) $\times$ U(1) $\times$ $\Theta$.
%Since $\Theta$ is discrete it is sometimes convenient to
%also just consider the group $\tilde G = $ SO(3) $\times$ U(1)
%(see below).
For $L=1$, it can be shown easily that all order parameters
can be written in the form $\hat k \cdot (\hat m + c \hat n) e^{i \gamma}$
where $\hat m$ and $\hat n$ are real orthogonal unit vectors,
and $c$ is a complex number with $|c| \le 1$.
(We shall, for simplicity of presentation, not always normalized
our wavefunction, since this normalization has no effect on
symmetries.  We can restrict ourselves to $|c| \le 1$ because
if otherwise, we can always divide the state by $c$ and
then rewrite the state in the above form).
It is also sufficient to consider only the two cases
$c=0$ and ${\rm Im} c \ne 0$
(if ${\rm Im} c = 0$, then the state is just 
$(\hat k \cdot \hat m' ) e^{i \gamma} $ for some $\hat m'$).
Special choices of $\hat m$, $\hat n$ and phase factor $\gamma$ results in
{\it reference} states, which we shall
take to be $ \hat k \cdot \hat z$ and
$\hat k \cdot (\hat x + c \hat y)$.  All states
$(\hat k \cdot \hat m) e^{i \gamma}$ can be
obtained from the reference state $\hat k \cdot \hat z$
by rotation and gauge transformation.
Equivalent statement applies to the states 
$\hat k \cdot (\hat m + c \hat n) e^{i \gamma}$
for a given $c$.  
[Technically, the collection of states $\{ g\Psi \}$ obtained from
$\Psi$ by operating the transformation $g$ on it are said
to form the {\em orbit} of $\Psi$].
As far as our problem of obtaining stationary states are 
concerned, these solutions related by symmetries
are "equivalent" solutions.  $\Psi$ is an inert state
if and only if $g \Psi$ is an inert state.

  An important consideration in determining whether a particular
state $\Psi$ is an inert state is the symmetry group under
which $\Psi$ is invariant, that is, the {\em isotropy subgroup} H
 of $\Psi$, where H contains all operations $h$ in G such
that $h \Psi = \Psi$.  We first note here that a state which
can be obtained by a transformation $g$ on $\Psi$ has
symmetry simply related to that of the original reference state $\Psi$.
[Technically, $g\Psi$ has the isotropy group
$gHg^{-1}$ thus these isotropy groups are said to be {\em conjugate}
to each other].
Consider now the reference state $\Psi_c = \hat k \cdot (\hat x + c \hat y)$.
If $c = c_1 + i c_2$ with $c_1 \ne 0$ and $c_2 \ne 0$
then H contains the identity $E$ alone. 
All states 
with $c_1 \ne 0$ and $c_2 \ne 0$ 
have the same symmetries.
[Technically, the collection of orbits 
$\{g \Psi_c\}$ from states $\Psi_c$ with the same symmetry
 are said to form a {\em stratum}].
 For any given
$(c_1,c_2) \ne (0,0)$, we can choose a $(c_1',c_2')$ arbitrary close
to the original $(c_1,c_2)$ such that $\Psi_{c'}$ 
($\ne g \Psi_c$ for any $g$ in G)
has the same symmetry as the original $\Psi_c$.
[Technically, the orbit $\{ g \Psi_c \}$ 
therefore {\em not isolated} in its stratum].
Michel's theorem then says that the state $\Psi_c$ is
not an inert state.
For example, we do not expect the state
$\hat k \cdot ( \hat x + \frac{1+i}{\sqrt{3}} \hat y)$
with $c = \frac{1+i}{\sqrt{3}}$ to be an inert state.  
While it may minimize a particular free energy $F$,
there is no reason to expect $c$ to remain at the
same value of $c = \frac{1+i}{\sqrt{3}}$ when the parameters
in the free energy $F$ are changed.
Similar remarks apply to the case where $c = i c_2$ with
$|c_2| < 1$ (though H contains also other elements in additional to E).

The states with, however, $c = i$ is special in that
the isotropy group H would then become  
$\{ R_z(\alpha) e^{- i \alpha}, U_2^{\gamma} e^{- 2i \gamma} \Theta \} $
where $R_z(\alpha)$ denotes a rotation of angle $\alpha$ about 
$\hat z$, and $U_2^{\gamma}$ denotes a $\pi$ rotation
about the horizontal axis $\hat \gamma = {\rm cos} \gamma \hat x
+ {\rm sin} \gamma \hat y$.
(Note that $\Theta$ does not commute with $e^{ - i 2 \gamma}$.
We are using the notation where operations start from the left).
A small modification of $c$ away from $i$ would
result in a state that has a completely different symmetry
(and is not $g \Psi$ for some $g$ in G, that is, 
not simply another order parameter related to $\Psi$ by
a transformation in G).
[Technically, we say that
the orbit $g \Psi$ is {\em isolated in its stratum}].
Michel's theorem then states that $\Psi_{c = i}$ is an inert
state.  This state is a stationary point of the free energy
irrespective of the parameters that enter in $F$.

Similar statement applies to $\Psi = \hat k \cdot \hat z$,
(related to the state $\Psi_{c=0}$ by a rotation)
where the isotropy group is H $= \{R_z(\alpha), U_2^{\gamma} e^{i \pi} \}
\times \Theta$.
(From the above argument, we have in fact shown that
the only inert states for L=1 are the axial state
$\hat k \cdot (\hat x + i \hat y)$ and the polar state
$\hat k \cdot \hat z$.  In the spin wavefunction language
they are $(1,0,0)$ and $(0,1,0)$ respectively, with
isotropy groups already given above).

Therefore, in general (not restricted to $L=1$),
one is searching for states $\Psi$ that have
special symmetries in the sense that a slight
deviation of the coefficients entering $\Psi$
would give another state  $\Psi'$
with an isotropy group different from H
(but barring cases where $\Psi'$ is just an
"equivalent" solution, that is, $\Psi' = g \Psi$
and thus just a symmetry-transformed $\Psi$,
i.e., cases where the new order parameter is in the same orbit
and thus with symmetry group $g{\rm H}g^{-1}$)
[Technically, we look for states $\Psi$ such
that its orbit is {\em isolated in its stratum}].

The procedure is thus to search for all subgroups H of
the combined group SO(3) $\times$ U(1) $\times \Theta$
with broken U(1) gauge symmetry which would determine
$\Psi$ uniquely.  Each of these $\Psi$ would then be
an inert state. (In principle it is sufficient that
H determines a set of possible $\Psi$'s, such
that the orbits of these order parameters are isolated.
However, in the present case of interest,
this ambiguity is only in an overall $\pm$ sign for the wavefunction.
Thus effectively each of these H determines a unique inert state
$\Psi$.)
It is useful to note that  once such an H is
found, then there is no need to consider the subgroups of H.
This is because any $\Psi_2$ invariant under a subgroup H$_2$ of 
H$_1$ but not H$_1$ itself cannot yield an inert state.  
(To see this, consider $\Psi(\lambda) = \Psi_2 + \lambda \Psi_1$
where $\Psi_1$ ($\ne \Psi_2$) is invariant under H$_1$. Then $\Psi(\lambda)$
is invariant under H$_2$ for any $\lambda$ and hence,
by the argument given, $\Psi_2$ cannot be an inert state.)

The subgroups of SO(3) $\times$ U(1) $\times \Theta$ 
that break the gauge symmetry can be constructed in
a similar manner as in \cite{VG}.  There, Volovik and Gorkov
constructed all possible isotropy subgroups for unconventional
superconductors where the normal state possesses discrete 
rotational symmetries of crystals instead of SO(3).
A principal observation, which still applies here,
is that U(1) is isomorphic to rotation about a fixed axis.
Thus the required groups would involve elements that
are in general product of a spatial rotation and a gauge transformation.

In our present case, these subgroups fall into two categories.
The first involves continuous groups.  These can be constructed 
easily (see below).  The second involves discrete groups.  Our task
is simplified since the discrete subgroups of 
SO(3) are well known.  They are the three groups of regular polyhedrons: 
O of the octahedron, T the tetrahedron, and Y 
the icosahedron; the dihedral groups D$_m$ and the cyclic groups $C_m$.

Before we actually list these groups we note further a great simplication.
Let us consider a general spin wavefunction
$\Psi = (\zeta_S, ... , \zeta_{-S})$.  If the group H contains
the symmetry element $h = C_n e^{i j 2 \pi/ n}$ where $j$ is
an integer which can be restricted to $-n < j < n$
($j$ is an integer follows from the requirement that $h^n$ must also be in H:
as $C_n^n=1$, we must have $j$ being an integer),
since under $C_n$, $\zeta_m \to \zeta_m e^{ - i m 2 \pi/n}$, we must
have $e^{ i (j-m) 2 \pi / n } = 1$ for $\zeta_m \ne 0$.
Hence only $\zeta_j$, $\zeta_{j \pm n} $ etc can be finite.
If there is only one such finite $\zeta_m$ then the order parameter
has been determined uniquely.
The state then has in fact continuous symmetry 
$R_z (\alpha) e^{i j \alpha}$.  If there are more than one non-zero $\zeta_m$,
say $\zeta_j$ and $\zeta_{j-n}$, then it can be an inert state only
if there are further other symmetry elements in H that puts a constraint
relating $\zeta_j$ and $\zeta_{j-n}$.  Only when $j-n = j$ can this be
achieved by a horizontal rotational axis, and the symmetry group would
then be a combined group constructed from the dihedral groups D$_n$
(or a larger group containing this).  If $j-n \ne j$, such requirements
would arise only from groups that are constructed from the 
regular-polyhedral groups O, T and Y.

Hence we can follow the following procedure.  First we consider all the 
continuous isotropy subgroups and find the corresponding order 
parameters.  
Then we consider the (non-dihedral) subgroups that
are constructed from the polyhedral groups:
O $\times$ U(1) $\times \Theta$,
T $\times$ U(1) $\times \Theta$, and Y $\times$ U(1) $\times \Theta$.
If an order parameter for a particular isotropy group can be found,
then we discard all its subgroups.  If not, then
we go to a  group of lower symmetry.  This is basically
the only part of the procedure that is more involved.
After such groups are exhausted, the only inert states
that can be left are those correspond to
the dihedral symmetries.  These states 
can be obtained by inspection (with the procedure already hinted above
in the last paragraph: see examples below).
For our case with $L \le 4$, we shall argue 
in Appendix \ref{Y} that five-fold symmetry elements
cannot be involved, hence there is no need to consider 
the icosahedral group Y (and also dihedral groups containing $C_5$).
The only non-dihedral polyhedral groups needed are thus those constructed from
O $\times$ U(1) $\times \Theta$ and
T $\times$ U(1) $\times \Theta$, and they can already be found
in \cite{VG}.

Now we proceed therefore to list the continuous subgroups, and 
subgroups which can be constructed from the polyhedral groups
O and T.  The continuous ones are easy. 
We can combine  $R_z(\alpha)$, continuous rotation of angle $\alpha$ 
about the axis $z$,
with gauge transformation $e^{ i n \alpha}$ where 
$n$ is an integer.  
The only state that would be invariant
under $R_z (\alpha) e^{ i n \alpha}$ for all $\alpha$
is the wavefunction where the only non-zero
component $\zeta_m$ in $\Psi$ is $\zeta_n$.
Corresponding wavefunction of angular momentum $L$ is
$\zeta_n Y_L^n (\hat k)$.
This wavefunction automatically has symmetry elements
around horizontal axes $\hat \gamma$ combined with suitable
phase factors (due to the definition from spherical harmonics).
For $L-n$ even, the isotropy group is 
\begin{equation}
{\rm H_n^e} \equiv \{ R_z (\alpha) e^{i n \alpha},
 U_2^{\gamma} e^{ 2 i n \gamma} \Theta \}
\label{une}
\end{equation}
whereas for $L-n$ odd, we have
\begin{equation}
{\rm H_n^o} \equiv \{ R_z (\alpha) e^{i n \alpha},
 U_2^{\gamma } e^{  i (2 n \gamma + \pi)} \Theta \}
\label{uno}
\end{equation}

We need only consider non-negative $n$'s,
since the state with only $\zeta_{-n}$ being non-zero
is related to the state with $\zeta_n$ being the only
non-zero element via either the time-reversal symmetry
or a horizontal rotation by $\pi$.

For the discrete (non-dihedral) subgroups contructed
from O and T, they have already been 
considered by Volovik and Gorkov \cite{VG}
and so we simply list their answers below:

\begin{equation}
{\rm O} \times \Theta \equiv \{ E, 8C_3, 3C_2, 6 U_2, 6C_4 \} \times \Theta
\label{O}
\end{equation}
This group corresponds to just breaking the gauge symmetry in 
O $\times$ U(1) $\times \Theta$.
Here $C_3$ are three-fold rotations along the diagonals, 
$U_2$ are two-fold rotations along the edges,
and $C_4$ are four-fold rotations about the faces of a cube,
and $C_2 = C_4^2$.

\begin{equation}
{\rm O(T)} \times \Theta \equiv
 \{ E, 8C_3, 3C_2, 6 U_2 e^{i \pi} , 6C_4 e^{i \pi} \} 
\times \Theta
\label{OT}
\end{equation}

This group is isomorphic to O, but contains only T as a subgroup.

\begin{eqnarray}
{\rm O(D_2)} &\equiv& 
\{ E, 3C_2, 2 C_4^x \Theta, 2 C_4^y \epsilon \Theta, 
    2 C_4^z \epsilon^2 \Theta, \nonumber \\
 & & 4 C_3 \epsilon, 4 C_3^2 \epsilon^2, 2 U_2^{\perp x} \Theta,
    2 U_2^{\perp y} \epsilon \Theta,  2 U_2^{\perp z} \epsilon^2 \Theta \}
\label{OD2}
\end{eqnarray}

This group contains $D_2$ as a subgroup.  Here $U_2^{\perp x}$ denotes
a $\pi$ rotation about the edge of a cube perpendicular to $x$
(i.e., $x \to -x$ and 
$y \leftrightarrow z$ or $y \leftrightarrow -z$),
and $\epsilon \equiv e^{2 i \pi/3}$.

We also have the tetrahedral groups

\begin{equation}
{\rm T} \times \Theta \equiv \{ E, 3 C_2, 4 C_3 , 4 C_3^2  \}
 \times \Theta 
\label{T}
\end{equation}

and

\begin{equation}
{\rm T(D_2)} \equiv \{ E, 3 C_2, 4 C_3 \epsilon, 4 C_3^2 \epsilon^2 \}
\label{TD2}
\end{equation}
This last group is isomorphic to T but only contains D$_2$ as
a subgroup.  We note here also that T is a subgroup of both O and O(T)
and ${\rm T(D_2)}$ is a subgroup of ${\rm O (D_2)}$.
It turns out that neither T nor ${\rm T(D_2)}$ 
give rise to inert states for any of the cases we have examined
below.

This complete the list of all non-dihedral groups that we would need
(for $L \le 4$).

\section{\label{results} Results}

Now we apply our theory to, in order $S$ (or $L$) $ = 2, 3, 4$.
For simplicity in notations, when writing the wavefunction
for $L$, we would simply write $x, y, z$ instead of 
$\hat k_x, \hat k_y, \hat k_z$.  It is understood that
$x^2 + y^2 + z^2 = 1$.

\subsection{$S = 2$}

The inert states for
$L = 2$ corresponding to d-wave Cooper pairing has in fact
been investigated before by \cite{Bruder86} (see also \cite{VW})
by considering the system as strongly spin-orbit coupled
$L=1$ and $S=1$ Cooper pairs with total angular momentum
$J = 2$.  Here we would like to present our arguments directly
in terms of the procedure outline above so that 
it can be easily compared with the results for $S = 3$ and $4$.

{\em Continuous Groups:}  We can easily write down the
three order parameters corresponding to continuous groups
${\rm H_n^{e,o}}$ for $n = 2, 1, 0$.  We shall call these
states F$_2$, F$_1$ and P$_0$
(Here F indicates {\em ferromagnetic} and P {\em polar},
using the language as in, e.g., \cite{Ciobanu00,HoYip99}).
Explicitly, the state wavefunctions are

\begin{eqnarray}
{\rm F_2} &\equiv&   (1, 0, 0, 0, 0)  \nonumber \\
{\rm F_1} &\equiv&   (0, 1, 0, 0, 0)  \nonumber \\
{\rm P_0} &\equiv&   (0, 0, 1, 0, 0)
\label{L2-con}
\end{eqnarray}
with corresponding polynomial forms
$(x+iy)^2$, $- z (x+ iy)$ and $ (2 z^2 - x^2 - y^2 )$.

{\em Discrete groups:} 
It can be easily verified that no order parameter
can satisfy the groups O $\times \Theta$, O(T) $\times \Theta$
and T $\times \Theta$.
To see this, we note that all $L=2$ wavefunctions can be
written as polynomials of  second degree in $x, y, z$.
The elements $3 C_2$ require $\Psi$ to be a linear combination
of $x^2$, $y^2$, $z^2$ only.
(e.g., $xy$ is odd under $C_2^x$ and hence must be rejected).
All these three groups contain eight three-fold rotations $C_3$ along
the diagonal of a cube, where $(x, y, z) \to (y, z, x)$, 
$(x,y,z) \to (-z, -x, y)$ etc.
The only function that has this property is
$x^2 + y^2 + z^2$, which however does not belong to $L=2$.

 The discrete group
O(D$_2$) in eq (\ref{OD2}) uniquely determines 
(up to a $\pm$ sign) the order parameter
to be $ x^2 + \epsilon y^2 + \epsilon^2 z^2$ (see also \cite{VG}).
We have just seen that $3 C_2$ limit us to 
the polynomials $x^2$, $y^2$ and $z^2$.
Under $C_3$, we have the mapping $( x y z) \to (y z x)$
and hence the elements $C_3  \epsilon$ requires
that the wavefunction be proportional to $x^2 + \epsilon y^2 + 
\epsilon^2 z^2$.   Elements such as $C_4^x \Theta$ 
fixes the overall phase factor to be real (under $C_4^x$,
$(x, y, z) \to (x, z, -y)$). We can also check that elements
such as $U_2^{\perp x} \Theta$ are satisfied.
As before \cite{Ciobanu00,HoYip99}, we shall refer to this state as the 
"cyclic" state. 
We shall denote it by C, and 
further discuss this state below.

Since we have already found the order parameter for O(D$_2$),
there is no need to consider its subgroup T(D$_2$).
This completes the non-dihedral polyhedral groups.

It remains then to consider the ones corresponding to
order parameters where only $\zeta_{\pm m} \ne 0$ for some
$m \ne 0$ and symmetries related to the dihedral groups.
  For $m=2$ we can have the state 
\begin{equation}
{\rm P_2} \equiv   ( 1, 0, 0, 0, 1)
\label{L2P2}
\end{equation}
Its polynomial form is $x^2 - y^2$ and easily seen to
possess the symmetry (same notation as in \cite{VG})

\begin{equation}
{\rm D_4 (D_2)} \times \Theta \equiv
\{ E, C_2, 2U_2, C_4 e^{ i \pi}, 2 U'_2 e^{ i \pi} \} \times \Theta
\label{D4D2}
\end{equation}

It can easily be checked that this group actually determines
the order parameter uniquely (again up to $\pm$ sign, a 
warning that we would not repeat again below) to be
$x^2 - y^2$, and hence the state is inert.
Note that a rotational about $z$ changes the relative
phase between $\zeta_{\pm 2}$, thus all states of
the form $ e^{ i \phi} ( e^{-2 i \alpha}, 0, 0, 0, e^{2 i \alpha})$
are related to ${\rm P_2}$  by operations within G.
States with $|\zeta_{-2}| \ne |\zeta_2|$ as the only non-zero
elements  have lower symmetries (in the sense of subgroup)
and therefore need not be considered.

One may attempt to try the state 
$\Psi = (0, 1, 0, 1, 0)$, which in polynomial form
reduces to $ - i 2 z y$.  Thus up to a gauge tranformation
and a rotation, this state is just the same as $2 x y$.
We note here that a $\pi/4$ rotation about $z$ changes
$x^2 - y^2$ to $2xy$.   Thus the state
$(0,1,0,1,0)$ can be transformed to the state P$_2$ and
does not correspond to a new inert state.

This completes our search for inert states for $L=2$.
We found five inert states:  F$_2$, $F_1$, P$_0$ with
continuous symmetries, C and P$_2$ with discrete symmetries.

In previous works \cite{Mermin74,Ciobanu00}, 
actual minimization of general quartic free energy
were performed.  Our inert states accounted for
all minina found in \cite{Mermin74,Ciobanu00}. 
There, it turns out that F$_1$ is never the absolute
minimum for any parameters entering $F[{\Psi}]$.
 It also turns
out to be a peculiar fact that, to this order,
P$_0$ and P$_2$ are always degenerate.  Furthermore,
any real linear combination between P$_0$ and P$_2$
are also degenerate \cite{Mermin74}
[Mermin has shown that, when the state is written in
the form $B_{ij} x_i x_j$ where $x_{1,2,3} = x, y, z$,
any real symmetric traceless $B_{ij}$ produces the same free energy.
Since any such matrix $B_{ij}$ can be diagonalized,
we see that these $\Psi$ can always be recast into
the form
$a ( 2 z^2 - x^2 - y^2 ) + b (x^2 - y^2)$ with $a$ and $b$ real.
This is thus a real linear combination of 
P$_0$ and P$_2$]
We however expect that these accidental degeneracies
would be lifted in more general free energies
(e.g. containing higher order terms in $\Psi$).

Let us further comment on the cyclic state.  
It can easily be checked that the polynomial
$ \epsilon x^2 + \epsilon^2 y^2 + z^2$ is
proportional to $ \sqrt{2} Y_2^0 + i (Y_2^2 + Y_2^{-2})$,
and hence the state vector can be written
as 
\begin{equation}
{\rm C} \equiv (i, 0, \sqrt{2}, 0, i)  
\label{L2C1}
\end{equation}
corresponding  to a 
form used in \cite{Ciobanu00}.  
A $\pi/4$ rotation around $\hat z$ can
transform this into $(1, 0, \sqrt{2}, 0, -1)$,
that is $\sqrt{2} Y_2^0 + (Y_2^2 - Y_2^{-2})$.
We remark here that
this state was in fact found by Anderson and Morel \cite{AM}
to be the minimum energy state 
for a d-wave Fermi superfluid
within weak-coupling BCS theory.
A pictorial way of showing the symmetry of this state
can also be found there (see also \cite{VG}).

Since the group O(D$_2$) has explicitly the symmetry
element $C_3 \epsilon$, we know immediately that,
if one uses quantization axis along this axis, the
wavefunction must assume the form
$( 0, \zeta_1, 0, 0, \zeta_{-2})$, {\it i.e.},
only $\zeta_1$ and $\zeta_{-2}$ are non-zero.
Indeed, let us take $\hat w \equiv ( \hat x + \hat y + \hat z) /\sqrt{3}$
and $\hat u \equiv ( - \hat x + \hat y) /\sqrt{2}$,
and $\hat v \equiv \hat w \times \hat u$ to be the new
orthonormal axes.  Using the transformation from 
$\vec r = x \hat x + y \hat y + z \hat z$ to
the new coordinates $\vec r = u \hat u + v \hat v + w \hat w$,
simple algebra shows that
$  \epsilon x^2 + \epsilon^2 y^2 + z^2 = 
 - \sqrt{2} i w (u + i v) - \frac{ (u - i v)^2 }{2}$.
Therefore, with $\hat w$ as the quantization axis,
the wavefunction can then be also written
as $(0, \sqrt{2} i, 0, 0, -1)$, the form as advertized.
Of course, there are also many other possible forms for C
(see., e.g., \cite{Ueda02}).
We remind the reader here that this state is 
unique up to gauge and rotations
(as also proven by Mermin \cite{Mermin74}).

\subsection{$S=3$}

The states that have continuous symmetries are

\begin{eqnarray}
{\rm F_3} &\equiv& (1, 0, 0, 0, 0, 0, 0) \nonumber \\
{\rm F_2} &\equiv& (0, 1, 0, 0, 0, 0, 0) \nonumber \\
{\rm F_1} &\equiv& (0, 0, 1, 0, 0, 0, 0) 
\label{L3c}  \\
{\rm P_0} &\equiv& (0, 0, 0, 1, 0, 0, 0) \nonumber
\end{eqnarray}
with corresponding polynomial forms
$ -(x + i y )^3$, $ z(x + i y)^2$,
$- (5z^2 -1) (x + i y)$, and $(5z^3 - 3 z)$.

To  construct the states belonging to the polyhedral groups,
it is convenient to first notice that an $L =3$ state
can always be written in the form of a cubic polynomial
in $x, y, z$, thus $B_{ijk} x_i x_j x_k$
where $x_{1,2,3} \equiv x, y, z$.  We can further
limit ourselves to $B_{ijk}$ symmetric under all
possible interchanges of indices $i j k$.
 Orthogonality
to $L=1$ then requires $ < B_{ijk} x_i x_j x_k x_l > = 0$
where $< ... > $ indicates an angular average.  Hence
we have $ \sum_i B_{lii} = 0$ and so
it is sufficient to specify only seven ($ = 2 \times 3 + 1$)
coefficients, which can be chosen to be
$B_{xyz}$, $B_{xyy}$, $B_{xzz}$, $B_{yxx}$, $B_{yzz}$, $B_{zxx}$,
$B_{zyy}$ (by eliminating e.g., $B_{xxx}$ in terms of $B_{xyy}$ and $B_{xzz}$.)
All the polyhedral groups O $\times \Theta$, O(T) $\times \Theta$,
O(D$_2$), ${\rm T} \times \Theta$ and T(D$_2$), 
contain $3 C_2$.  The only polynomial
that satisfies this is $xyz$.
(e.g., invariance under $C_z^x$ leaves only $xyz$, $xyy$ and $xzz$.
The latter two are eliminated by the requirement of invariance
under $C_2^z$.)  Therefore obviously O, O(D$_2$), T(D$_2$)
cannot be satisfied.  However, it does satisfy O(T) $\times \Theta$.
We can check in fact O(T) $\times \Theta$ uniquely determines
the order parameter to be $xyz$.  Note that $x y z = z {\rm Im} (x+iy)^2$,
hence the spin-wavefunction for $xyz$ can be written
as $(0, -i, 0, 0, 0, i, 0)$. 
Though by a suitable rotation about $z$, we can change 
this to the form $(0,1,0,0,0,1,0)$ and thus call this ${\rm P_2}$,
for reasons to be explained below
we shall instead call this state D:

\begin{equation}
{\rm D} \equiv (0, -i, 0, 0, 0, i, 0)
\label{L3P2}
\end{equation}

Since T $\times \Theta$ is a subgroup of O(T) $\times \Theta$,
we have exhausted all states under the non-dihedral polyhedral groups.
Hence we proceed to the states with only $\zeta_{\pm m}$ being finite
obeying dihedral groups.
For $m=3$ we can have the state $(-1,0,0,0,0,0,1)$, and 
we shall call this P$_3$.  It has the polynomial form
${\rm Re} (x + i y)^3$ and the isotropy group
\begin{equation}
{\rm D_6(D_3)} \times \Theta \equiv
\{ E, 2 C_3, 3 U_2, C_2 e^{i \pi}, 2 C_6 e^{i \pi}, 3 U'_2 e^{ i \pi} \}
\times \Theta   \ .
\label{D6D3}
\end{equation}
We can check again that ${\rm D_6(D_3)}$ uniquely determines
the order parameter to be ${\rm Re} (x + i y)^3$.
For $m=2$, we would obtain the state P$_2$ which was already
discussed under the group O(T) $\times \Theta$.
For $m=1$, the wavefunction 
$(0,0,0,1,0,1,0,0,0)$
would then have the form $(5 z^2 -1) y$
up to a phase factor $i$.  This state is equivalent to
$(5 z^2 -1) x$ which has the symmetry
${\rm D_2 (U_2)} \times \Theta =
\{ E, C_2^z e^{ i \pi}, U_2, U'_2 e^{ i \pi} \} \times \Theta $ and is thus
a subgroup of ${\rm D_6 (D_3)} \times \Theta$ considered before and
thus this state is not an inert state. 
We thus have exhausted the list for states with dihedral symmetries.

Summarizing,  for $S=3$ we found six inert states,
F$_3$, F$_2$, F$_1$ and P$_0$ with continuous symmetries,
D and P$_3$ with discrete symmetries.  

Minimization of a general quartic free energy has been
carried out by  \cite{Diener06,Santos06}.  In \cite{Diener06},
F$_3$, F$_2$ and P$_3$ appeared explicitly as absolute minima
for suitable interaction parameters.  They also find 
a state (which they named state D) which we shall call ${\rm D'}$
with the spin wavefunction
$(\sqrt{2}, 0, 0, i \sqrt{5}, 0, 0, \sqrt{2})$.
The form suggests that it is an inert state.  Since it has
three-fold symmetry but it is 
obviously not P$_3$, this suggests that it should be the
state D expressed in rotated axis.  We shall verify that 
this is indeed the case in Appendix \ref{D}.  Thus their state
${\rm D'}$ has isotropy group (conjugate to) O(T) $\times \Theta$.

\subsection{$S=4$}

For continuous groups, we have
\begin{eqnarray}
{\rm F_4} &\equiv& (1, 0, 0, 0, 0, 0, 0, 0, 0) \nonumber \\
{\rm F_3} &\equiv& (0, 1, 0, 0, 0, 0, 0, 0, 0) \nonumber \\
{\rm F_2} &\equiv& (0, 0, 1, 0, 0, 0, 0, 0, 0) 
\label{L4c} \\
{\rm F_1} &\equiv& (0, 0, 0, 1, 0, 0, 0, 0, 0) \nonumber \\
{\rm P_0} &\equiv& (0, 0, 0, 0, 1, 0, 0, 0, 0) \nonumber \\
\end{eqnarray}
with polynomial forms $(x + iy)^4$,
$ - z (x+iy)^3$, $(7z^2-1)(x+iy)^2$, $- (7z^3-3z)(x+iy)$,
$(35z^4 - 30z^2 + 3)$.
For discrete groups, we have to consider even order polynomials
up to fourth order (we shall require orthogonality to $L=0$ and $2$ later).
For the polyhedral groups O $\times \Theta$, O(T) $\times \Theta$,
O(D$_2$), T $\times \Theta$ and T(D$_2$), the elements $3 C_2$
eliminated all polynomials except
$x^4, y^4, z^4, y^2z^2, z^2x^2, x^2y^2, x^2, y^2, z^2$.
To obey $8 C_3$ in O $\times \Theta$ and O(T) $\times \Theta$,
we must then have the linear combination
\begin{equation}
\Psi = c (x^4 + y^4 + z^4) + a (y^2z^2 + z^2 x^2 + x^2 y^2) 
  + b(x^2 + y^2 + z^2)
\label{trial1}
\end{equation}
 Obviously O(T) $\times \Theta$
is not satisfied (no required sign changes under $C_4$),
 but O $\times \Theta$ is if 
$c, a, b$ are all real.
Before we continue, we note that the three functions
multiplying the coefficients $c, a, b$ are not linearly
independent.  Indeed, using $x^2+y^2+z^2=1$
we can verify that
$(x^4 + y^4 + z^4) = 
  (x^2 + y^2 + z^2) - 2 (y^2z^2 + z^2 x^2 + x^2 y^2) $.
Let us choose to use 
$(y^2z^2 + z^2 x^2 + x^2 y^2)$ and $x^2 + y^2 + z^2 = 1$
to be the two independent functions, thus setting $c=0$ in 
eq (\ref{trial1}).   
  Orthogonality to $L=0$ and $2$
requires $<\Psi> = 0$ , $<\Psi x^2> =  
<\Psi y^2> = <\Psi z^2> =0$.
Evaluating the angular averages \cite{note}, we find that
{\em all} these relations are satisfied if $b = -  a / 5$.
(The fact that these equations are dependent is not a coincidence, since
$< x^2 (x^4 + y^4 + z^4) >/ < (x^4 + y^4 + z^4) >  =$
$< x^2 (y^2z^2 + z^2x^2 + x^2y^2) > / <(y^2z^2 + z^2x^2 + x^2y^2) >$
$= < x^2 > =  1/3$ etc from symmetry.)
Thus ${\rm O} \times \Theta$  determines the order
parameter uniquely and the state is inert.
Choosing $a = 5$ gives us the wavefunction
\begin{equation}
\Psi = 5 (y^2z^2 + z^2 x^2 + x^2 y^2) -1
\label{L4O}
\end{equation}
For lack of a better name we shall just call this state ${\rm O}$.
In Appendix \ref{C} we shall verify that the corresponding
spin wavefunction, up to a normalization factor and sign, is given by
\begin{equation}
{\rm O} \equiv (\sqrt{5}, 0, 0, 0, \sqrt{14}, 0, 0, 0, \sqrt{5}) \ .
\label{PsiO4}
\end{equation}

We need not consider ${\rm T} \times \Theta$ since it is 
a subgroup of ${\rm O} \times \Theta$.

In O(D$_2$) or T(D$_2$), the element $C_3 \epsilon$
require that we have instead the linear combination
\begin{eqnarray}
\Psi &=& c (x^4 + \epsilon y^4 + \epsilon^2 z^4)  \nonumber \\
  & & +  a (y^2z^2 + \epsilon z^2 x^2 + \epsilon^2 x^2 y^2) 
  + b(x^2 + \epsilon y^2 + \epsilon^2 z^2)
\label{trial2}
\end{eqnarray}
So far, $c,a,b$ can be complex.  This state obviously satisfies
${\rm T(D_2)}$ (after orthogonality to $L=0$ and $2$ have been shown).
To satisfy the larger group ${\rm O(D_2)}$, $C_4^x \Theta$
requires that the coefficients $c, a, b$ are all real.
[In this case, one can check that elements such as $U_2^{\perp x} \Theta$
($x \leftrightarrow -x$, $y \leftrightarrow z$ and then complex conjugate)
are also satisfied.]  
Before we proceed, we note again that the three functions
multiplying $c, a, b$ in eq (\ref{trial2}) are not linearly independent.
Indeed, using $x^2 + y^2 + z^2 = 1$ and $1 + \epsilon + \epsilon^2 = 0$,
we can verify that
$(x^4 + \epsilon y^4 + \epsilon^2 z^4) =
   (x^2 + \epsilon y^2 + \epsilon^2 z^2)
 + (y^2z^2 + \epsilon z^2 x^2 + \epsilon^2 x^2 y^2) $.
Let us proceed to work with the form with $c=0$.
It remains to find 
$a, b$ to satisfy orthogonality to $L = 0$ and $2$.
$<\Psi> = 0$ is trivial (since $1 + \epsilon + \epsilon^2 = 0$).
$<\Psi x^2 > = 0$  and $<\Psi y^2 > = 0$ 
both require $ - a + 7b = 0$
($<\Psi z^2 > = 0$ is then automatically satisfied),
and therefore both $a$ and $b$ can be chosen real. 
Hence we obtain
\begin{equation}
\Psi = 7 (y^2z^2 + \epsilon z^2 x^2 + \epsilon^2 x^2 y^2) 
  + (x^2 + \epsilon y^2 + \epsilon^2 z^2)
\label{L4OD2}
\end{equation}
as a state satisfying O(D$_2$) (thus there is no need to consider
${\rm T(D_2)}$).  In appendix \ref{C}, we show that, up to 
a renormalization constant and phase factor, 
the corresponding spin wavefunction
is 
\begin{equation}
{\rm C} \equiv 
(\sqrt{7}, 0, \sqrt{12}i, 0, -\sqrt{10},0, \sqrt{12}i, 0, \sqrt{7})
\label{PsiC4}
\end{equation}
This is our state with O(D$_2$) symmetry.  We are calling
this state cyclic (C) in analogy with the cyclic state in $L=2$.
This completes our consideration of polyhedral groups.

The states with dihedral groups can be found as in $L=2$ and $3$.
The state ${\rm P_4} \equiv (1,0,0,0,0,0,0,0,1)$
has polynomial form ${\rm Re} (x+iy)^4$ and symmetry
\begin{equation}
{\rm D_8(D_4)}\times \Theta
\equiv \{ E, C_2, 2C_4, 4 C_8 e^{i\pi}, 4 U_2, 4U''_2 e^{i \pi} \}
\times \Theta
\end{equation}
The state ${\rm P_3} \equiv  (0, i, 0, 0, 0, 0, 0, -i, 0)$,
(related to $(0,1,0,0,0,0,0,1,0)$ by a rotation around $z$),
polynomial form $z {\rm Im} (x+iy)^3$, has
symmetry ${\rm D_6(D_3)} \times \Theta$ (eq (\ref{D6D3})).
We need not consider $(0,0,1,0,0,0,1,0,0)$ since then it
has the form $(7z^2 - 1) (x^2 - y^2)$ and symmetry
${\rm D_4(D_2)} \times \Theta$, and thus is only a subgroup
of ${\rm D_8(D_4)} \times \Theta$.  Also,
for the similar reason there is no need to consider
$(0,0,0,1,0,1,0,0,0)$ which is proportional to $(7z^3 - 3z) y$
which has symmetry ${\rm D_2(C_2)}\times \Theta$.
This completes the list of discrete groups.

Summarizing, we found nine inert states:
${\rm F_4, F_3, F_2, F_1, P_0}$ with continuous symmetries
and ${\rm O, C, P_4, P_3}$ with discrete symmetries.

\section{\label{vortices} Vortices}

As an application of our results, we consider topological
line defects (vortices) of the spin condensates.
The distinct types of vortices have already been considered
in \cite{Ho98} for $S=1$ and \cite{Makela03} for the states
${\rm F_2, F_1}$ and C of $S=2$.  

The possible types of vortices are determined by the 
map of a circle in real space to the space of non-equivalent
order parameters.
  We apply the theorem that has been proven in, e.g,
\cite{Mermin79}, which states that 
$\pi_1 ({\rm G'/H'}) = {\rm H'/H'_0}$.  Here, ${\rm G'}$
is a simply-connected group characterizing the 
symmetry of the fully symmetric (normal) phase,
and ${\rm H'}$ is the isotropy subgroup 
(within ${\rm G'}$) of the state under
consideration, and ${\rm H'_0}$ is the part of ${\rm H'}$ that is
connected (in the sense of a neighborhood) 
with the identity $E$.  If ${\rm H'/H'_0}$ is
abelian, the line defects are simply classified by the elements
of this group.  Otherwise, the topologically distinct
type of vortices are classified by the conjugate classes of this group.

\subsection{S=2}

This has already been treated in \cite{Makela03}
({\it c.f.} \cite{Barnett06}) except the states
${\rm P_2}$ and ${\rm P_0}$.  For ease of comparison with
$S=3$ and $4$ however we shall also repeat the results for the other
states below.

For the states ${\rm F_{1,2}}$, since a rotation about the z axis
would also generate a gauge transformation, it is simpliest \cite{Mermin79}
to just consider the symmetry group SO(3) for the normal state.
For ${\rm F_1}$, a reference state can be taken just as
$z(x+iy)$.  Then H contains the identity $E$ alone.  To have
a simply-connected $\rm G'$, we "lift" G = SO(3) to $\rm G'=SU(2)$, and then
$\rm H'$ becomes ${\rm H'} =\{E,Q\}$
where $Q \ne E$ and $Q^2 = E$. ${\rm H'_0}$ 
consists of $E$ alone and ${\rm H'/H'_0 = H'}$.
 We thus have two classes
of vortices for ${\rm F_1}$.  

For ${\rm F_2}$, we take the reference state $(x+iy)^2$.
Then H = $\{ E, R_z(\pi) \}$ 
consisting of two classes.  When we lift from
G = SO(3) to $\rm G' = SU(2)$,  each element in H is doubled.  Thus
there are four classes of vortices.

For ${\rm P_0}$, we choose the reference state 
$(2 z^2 - x^2 - y^2)$ and take G = SO(3) $\times$ U(1).
Then ${\rm H = H_0^e} = \{  R_z(\alpha), U_2^{\gamma} \}$ (see eq (\ref{une})
consisting of two classes.  We need however a simply-connected $\rm G'$,
which can be taken to be ${\rm G'= SU(2)} \times T_{\phi}$,
where $T_{\phi}$ is the translational group for the phase $\phi$.
Then ${\rm H'} = \{ u(\hat z, \alpha) T_{2 n \pi}, 
  u(\hat \gamma, \pm \pi) T_{2 n \pi} \}$.
Here, we have used the representation 
$u(\hat n, \alpha) = {\rm cos} \frac{\alpha}{2} 
  + i \hat n \cdot \sigma {\rm sin} \frac{\alpha}{2} $
in terms of Pauli matrices for the rotations in SU(2).
${\rm H'_0} = \{ u(\hat z, \alpha) \}$, and thus
${\rm H'/H'_0}$ consists of the elements (cosets)
$ ... $, $\{ u(\hat z, \alpha) T_{- 2 \pi} \}$, 
$\{ u(\hat z, \alpha) \}$, $\{ u(\hat z, \alpha) T_{+ 2 \pi} \}$, 
$ ...$, $\{ u (\hat \gamma, \pm \pi) T_{2 \pi} \} $, $...$.
Since a product between $u(\hat z, \alpha_1)$ and $u(\hat z, \alpha_2)$
or between $u(\hat \gamma_1, \pm \pi)$ and $u(\hat \gamma_2, \pm \pi)$
both yield $u(\hat z, \alpha)$ for some $\alpha$;
and the product between $u(\hat z)$ and $ u(\hat \gamma, \pm \pi)$ 
yield another
element $u (\hat \gamma', \mp \pi)$,  
$\rm H'/H'_0$ is isomorphic to ${\rm Z_2 \times Z}$,
where we can assign the integer $0$ to $\{ u(\hat z, \alpha) \}$
and $1$ to $\{ u(\hat \gamma, \pm \pi) \}$ for the group ${\rm Z_2}$
and the circulation number $n$ for the group ${\rm Z}$.
Note this is different \cite{Makela03} from the polar state ${\rm P_0}$ for spin
$S=1$.  The source of difference is that the isotropy group
for the state ${\rm P_0}$ is ${\rm H_0^e}$ for $S=2$ but
${\rm H_0^o}$ for $S=1$.

The cyclic state C has been treated in \cite{Makela03} using
the language of the spin-wavefunction and its rotational properties.
Here the symmetries we obtained before becomes handy.
If ${\rm G = SO(3) \times U(1) \times \Theta }$,
we have already found ${\rm H = O(D_2)}$ (eq (\ref{OD2}).
To consider the vortices, we only need to consider instead
${\rm G = SO(3) \times U(1) }$, and then we obtain
${\rm H} = \{ E, 3C_2, 4 C_3 \epsilon, 4 C_3^2 \epsilon^2 \} $.
This group is isomorphic to the tetrahedral group with
${\rm  T} = \{ E, 3C_2, 4 C_3, 4 C_3^2 \} $ which has
four conjugate classes.  The double group $\rm T'$ corresponding T
has seven classes \cite{LL},
$\{E\}, \{Q\}, \{3C_2, 3C_2 Q \}, 
\{4C_3\}, \{4C_3Q\}, \{4C_3^2\}, \{4 C_3^2 Q \}$,
as each class in T is
doubled except the one ($\{3C_2\}$) which consists of 
bilateral two-fold axes \cite{Mirman}.
When G = SO(3) $\times$ U(1) is lifted to
the ${\rm G' = SU(2)} \times  T_{\phi}$,
we get ${\rm H'_0} = \{E\}$, ${\rm H'/H'_0 = H' }$ where
the group H' has classes
$\{ E T_{2 n \pi} \}$, $\{Q T_{2 n \pi} \}$,
$\{ 3 C_2 T_{2 n \pi}, 3 C_2 Q T_{2 n \pi} \}$,
$\{ 4 C_3 T_{ (2 n + \frac{2}{3}) \pi} \}$,
$\{ 4 C_3 Q T_{ (2 n + \frac{2}{3}) \pi} \}$,
$\{ 4 C_3^2 T_{ (2 n + \frac{4}{3}) \pi} \}$,
$\{ 4 C_3^2 Q T_{ (2 n + \frac{4}{3}) \pi} \}$,
where $n$ is an integer.  Thus the vortices are divided
into seven types in additional to the circulation number $n$.
Note however due to the presence of $\epsilon = e^{ i 2 \pi /3}$
factors in H and hence ${\rm H'}$, ${\rm H'/H'_0 }$ is not isomorphic
to the product of the double group $\rm T'$ times the discrete
translational group $T_{2 n\pi}$.

The polar state ${\rm P_2}$ with discrete symmetry can be treated similarly. 
 If G = SO(3) $\times$ U(1), then H is simply
${\rm D_4(D_2)}$ $= \{ E, C_2, 2U_2, 2C_4 e^{ i \pi}, 2U'_2 e^{ i \pi} \}$
isomorphic to ${\rm D_4 } = \{ E, 2C_2, 2U_2, 2C_4, 2U'_2 \}$
with five classes.  The double group ${\rm D'_4 }$ corresponding
to ${\rm D_4}$ consists of seven classes \cite{LL}:
$\{E\}, \{Q\},  \{C_2, C_2Q \}, \{2U_2, 2U_2 Q\},
\{ C_4, C_4^3 Q \}, \{C_4^3, C_4 Q \}, \{ 2U'_2, 2U'_2 Q \}$.
When we lift to the connected group ${\rm G' = SU(2)} \times T_{\phi} $,
we obtain ${\rm H'_0} = \{E\}$,
and ${\rm H'/H'_0 = H'}$ consists of the following classes:
$\{ E T_{2 n \pi} \}$, $\{Q T_{2 n \pi} \}$,
$\{  C_2 T_{2 n \pi},  C_2 Q T_{2 n \pi} \}$,
$\{ 2 U_2 T_{2 n \pi}, 2 U_2 Q T_{2 n \pi} \}$,
$\{  C_4 T_{2 n \pi},  C_4^3 Q T_{2 n \pi} \}$,
$\{  C_4^3 T_{(2 n +1)  \pi},  C_4 Q T_{(2 n +1) \pi} \}$,
$\{ 2 U'_2 T_{(2 n +1) \pi}, 2 U'_2 Q T_{(2 n+1) \pi} \}$.
There are seven vortex types in additional to the circulation numbers $n$.

\subsection{$S=3$}

For the state ${\rm F_3}$, it is again simpliest just to use
SO(3) and the reference state $(x+iy)^3$.  H consists of
three elements, $E, R_z(2 \pi/3), R_z(4 \pi/3)$.
The corresponding double group $\rm H'$ has six elements.
There are six distinct classes of vortices.

${\rm F_2}$ and ${\rm F_1}$ has the same corresponding H
as their $S=2$ counterpart.  The distinct classes of vortices
are again four and two respectively.

For the polar state ${\rm P_0}$, we choose the reference state 
$(5 z^3 - 3 z)$ and take G = SO(3) $\times$ U(1).
Then H =  $\{ R_z(\alpha), U_2^{\gamma} e^{i \pi}\}$  (see eq (\ref{uno}))
consisting of two classes. When using the
connected ${\rm G'= SU(2) }\times T_{\phi}$,
${\rm H'} = \{ u(\hat z, \alpha) T_{2 n \pi}, 
  u(\hat \gamma, \pm \pi) T_{(2 n +1) \pi} \}$.
${\rm H'_0} = \{ u(\hat z, \alpha) \}$, and thus
${\rm H'/H'_0}$ consists of the elements (cosets)
$ ... $, $\{ u(\hat z, \alpha) T_{- 2 \pi} \}$, 
$\{ u(\hat z, \alpha) \}$, $\{ u(\hat z, \alpha) T_{+ 2 \pi} \}$, 
$ ...$, $\{ u (\hat \gamma, \pm \pi) T_{ \pi} \} $, $...$,
Note now that the cosets with $u(\hat \gamma, \pm \pi)$ 
are associated with translation of phase of odd integral multiple
of $\pi$.  Again noting the properties of the products of
$u(\hat n, \alpha)$ discussed in $S=2$, 
we find that $\rm H'/H'_0$ is isomorphic to ${\rm Z}$,
as it is the case \cite{Makela03} for the state ${\rm P_0}$ belong to $S=1$.

For the state ${\rm D}$, when we just use 
${\rm G = SO(3) \times U(1)}$ then
${\rm H = O(T)} = \{ E, 8C_3, 3C_2, 6U_2 e^{i \pi}, 6C_4 e^{i \pi} \}$
is isomorphic to $O =  \{ E, 8C_3, 3C_2, 6U_2, 6C_4 \}$
with thus five classes.
The double group $\rm O'$ corresponding to O has eight classes,
(only $\{E\}$ , $\{8C_3\}$ and $\{6C_4\}$ are doubled).
Vortices can therefore be classified into the following classes:
$\{E T_{2 n \pi} \}$, $\{Q T_{2 n \pi} \}$,
$\{ 4C_3, 4C_3^2 Q \} T_{2 n \pi} $,
$\{ 4C_3^2, 4C_3 Q \} T_{2 n \pi} $,
$\{ 3C_2, 3C_2 Q \} T_{2 n \pi} $,
$\{ 3C_4, 3C_4^3 Q \} T_{(2 n+1) \pi} $,
$\{ 3C_4^3, 3C_4 Q \} T_{(2 n+1) \pi} $,
$\{ 6U_2, 6U_2 Q \} T_{(2 n+1) \pi} $.
There are thus eight classes in additional to the circulation
number $n$, and those elements originally associated
with $e^{i\pi}$ in the group O(T) are now associated with
phase translations $(2n+1)\pi$.

For the polar state ${\rm P_3}$, if we just use
${\rm G = SO(3) \times U(1)}$ then 
${\rm H = D_6(D_3)} =  
\{E, C_2, 2C_3, 2C_6 e^{i \pi} , 3U_2, 3U'_2 e^{ i \pi} \}$
was already given in eq (\ref{D6D3}).  This group
is isomorphic to ${\rm D_6} = 
\{E, C_2, 2C_3, 2C_6, 3U_2, 3U'_2 \}$ with six classes.  
The double group ${\rm D'_6}$ corresponding to ${\rm D_6}$ 
has nine classes:  all classes are doubled except
$C_2$, $3U_2$ and $3U'_2$.  Vortices are therefore divided to nine
classes, in additional to an integer $n$ specifying the circulation.
(Those classes that correspond to $C_6 e^{i\pi}$
and $U'_2 e^{i \pi}$ are associated with phase translations
$T_{(2n+1)\pi}$ rather than $T_{2n\pi}$
(c.f. the state D).

\subsection{$S=4$}
These can be obtained similarly and so we shall be brief.
${\rm F_1, F_2, F_3, F_4}$ have respectively
two, four, six and eight classes of vortices.
${\rm P_0}$ has symmetry same as ${\rm P_0}$ of $S=2$ and thus with the same
classes of vortices.  
A parallel statement applies to the cyclic state C.
The state O has eight classes of vortices in additional
to the circulation numbers.   We note however
that all vortices for O have winding phase factors $2 n \pi$,
thus {\em different} from the state D of $S=3$
where the symmetry group is O(T) instead of O.
The state ${\rm P_4}$ has symmetry
group ${\rm D_8 (D_4)}$.  
The double group ${\rm D'_8}$ has nine classes.
So vortices for ${\rm P_4}$ has nine classes of vortices
in additional to the circulation number, and each
class is associated phase translations of $(2n+1) \pi$ or $2n \pi$
depending on whether the factor $e^{i\pi}$ appears or
not in ${\rm D_8 (D_4)}$.
The state ${\rm P_3}$ is analogus to that in $S=3$.

\section{\label{conclusions} Conclusions}

To conclude, we have listed all inert states for a spin
superfluid for $S \le 4$.  These states are always
stationary points of any general free energies.
We have also given the residual symmetries 
(hence broken symmetries) of these
inert states.

This research is supported by the
National Science Council of Taiwan under Grant No.
 NSC95-2112-M-001-054.

\appendix

\section{\label{Y} Groups with five-fold symmetry elements}

Here we explain why groups containing five-fold symmetry elements
$C_5$ (without containing the continuous symmetry C$_{\infty}$), in particular 
the icosahedral group Y,  are not involved for inert states with
$L \le 4$.  
First, it is clear that, since $L \le 4$, the wavefunction
cannot be invariant under $C_5$ itself.  
Next we show that H cannot contain $\Theta$ itself.
For if it did, then it would contain both 
$C_5 e^{i \phi_5}$ and $C_5 e^{i \phi_5} \Theta$ for some
$\phi_5$.  The latter element can be rewritten
as $\Theta C_5 e^{ - i \phi_5}$ and hence H would contain
$C_5^2$, leading to a contradiction.
Similarly we can show that H cannot contain
an element of the form $h = C_5 e^{i \phi_5} \Theta$.
If $ h $ is an element in H,
then so must $h^5 = e^{i \phi_5} \Theta$.  Thus apart
from a gauge transformation, the state $\Psi$ is time-reversal
invariant ($\Psi' = \Psi e^{ i \phi_5/2}$ is invariant under $\Theta$).
which is again not possible.
Therefore $C_5$ must be in combination with a gauge transformation
only.  As seen
before, this phase factor must be integral multiple of $ 2 \pi / 5$. 
Hence let us 
consider the element $C_5 e^{ i j 2 \pi / 5}$ where $j$ is
an integer.  As seen already the non-zero elements of $\zeta_m$ must
be $\zeta_j$, $\zeta_{j-5}$ etc, and to have discrete but not continuous
symmetry at least two of such elements must be non-zero.
For $L = 1$ and $2$ this is not possible.  For $L = 3$,
the only possible pair is $(\zeta_3, \zeta_{-2})$ if $j=3$,
(equivalently $(\zeta_{2}, \zeta_{-3})$), and for $L = 4$
an additional possibility $(\zeta_4, \zeta_{-1})$ if $j=4$.
For these states to 
be inert one must have other symmetry operations in H that
would impose a relation between these non-zero elements.
Clearly a horizontal rotation (and hence groups like
D$_5$) is not sufficient, and hence one
is left with the icosahedral group Y. 
We recall \cite{QC} first that the icosahedral group Y contains,
in additional to the five-fold axis pointing along the 
(12) vertices, three-fold axes perpendicular
to the (20) triangular faces and (15) two-fold axes pointing
out to the (30) edges of the icosahedron.
 Recall also that in the group Y,  we have
$C_5 C_3 = C_2$, where $C_3$
is a three-fold axis normal to the triangle which
contains the $C_5$ as an vertex,
and $C_2$ is around the edge of the triangle
connected to the mentioned vertex for the $C_5$ axis \cite{QC}.
In the isotropy group H, these
two- and three-fold rotations $C_2$ and $C_3$
must appear as symmetry elements, though they can be combined with 
again phase factors and/or time reversal symmetry.
By similar argument as given already for $C_5$, $C_3$ cannot be
appear in the form $C_3 e^{ i \phi_3} \Theta$
(else the state is again time-reversal invariant),
thus the three-fold rotations can appear only via $C_3 e^{ i j' 2 \pi / 3}$
for some integer $j'$.  Therefore now H must contain the element
$C_5 e^{ i j 2 \pi / 5} C_3 e^{ i j' 2 \pi / 3}
= C_2 e^{ i ( \frac{j}{5} + \frac{j'}{3}) 2 \pi}  \equiv h_2$
and hence $h_2^2$.  But this is possible only if $j = j' = 0$
since $C_2^2 = 1$.
Thus we arrive again at the conclusion that H must
contain $C_5$ itself, which however is impossible for $L \le 4$.
This completes our proof.

\section{\label{D} Relation between states D and ${\rm D'}$}

We here show that the state $xyz$, with symmetry
O(T) $\times \Theta$ (eq (\ref{OT})),  is related to
\begin{equation}
{\rm D'} \equiv (\sqrt{2}, 0, 0, i \sqrt{5}, 0, 0, \sqrt{2})
\end{equation}
in a new set of orthonormal axes $(\hat u, \hat v, \hat w)$.
Since $\rm D'$ has explicitly three-fold symmetry along its
quantization axis (identified as $\hat w$) and
two-fold axes (up to $e^{i \pi}$) perpendicular to it,
and $xyz$ has
a three-fold axis along the cube diagonal, 
this suggests that we take 
$\hat w = (\hat x + \hat y + \hat z) / \sqrt{3}$,
$\hat u = (-\hat x + \hat y) /\sqrt{2}$ along
one of the $U_2 e^{i \pi}$ axis, 
and $\hat v = \hat w \times \hat u$
completing the triad.
This transformation between $x, y, z$ and $u, v, w$ is
the same as the one used in the $S=2$ cyclic state.
We have the relations
$x = - \frac{1}{\sqrt{2}} u - \frac{1}{\sqrt{6}} v + \frac{1}{\sqrt{3}} w$,
$y =  \frac{1}{\sqrt{2}} u - \frac{1}{\sqrt{6}} v + \frac{1}{\sqrt{3}} w$,
and $z = \sqrt{\frac{2}{3}}v + \frac{1}{\sqrt{3}} w$, 
therefore 
\begin{displaymath}
xyz = - \sqrt{2} v ( 3 u^2 - v^2) + w ( 2 w^2 - 3 u^2 - 3 v^2)   \ .
\end{displaymath}
On the other hand, we have (in the $u, v, w$ axis)
$Y_{3}^{\pm 3} = \mp \sqrt{\frac{35}{\pi}}\frac{1}{8} (u \pm i v)^3$,
$Y_{3}^{0} = \sqrt{\frac{7}{\pi}}\frac{1}{4} (5 w^2 - 3)$,
so state ${\rm D'}$ is proportional to
\begin{displaymath}
\sqrt{\frac{35}{\pi}}\frac{i}{4} 
[ - \sqrt{2} v ( 3 u^2 - v^2) + w ( 5 w^2 - 3 )] \ .
\end{displaymath}
Using $u^2 + v^2 + w^2 =1 $, we see that
state ${\rm D'}$ is proportional to $i$ times the state $xyz$.
We conclude that ${\rm D'}$ in $(u,v,w)$ axes is the same
as $ i {\rm D} = (0,1,0,0,0,-1,0) $ in $(x,y,z)$ axes.

\section{\label{C} Spin wavefunctions for states O and C of $S=4$}

We would like to find the spin-wavefunction corresponding
to states O  [eq (\ref{L4O})] and C [eq (\ref{L4OD2})].  
We note that both these expressions 
 contain only even powers of $x$, $y$ and $z$,
they must be just linear combinations of 
$Y_4^4+Y_4^{-4}$, $Y_4^2+Y_4^{-2}$ and $Y_4^0$.  With $c_4 \equiv 
\sqrt{\frac{9}{4 \pi}}$, we have
$Y_4^4+Y_4^{-4} = c_4 \sqrt{\frac{35}{32}}
 \left((x^2+y^2)^2-8x^2y^2\right)$, 
$Y_4^2+Y_4^{-2} = c_4 \sqrt{\frac{5}{8}} (7z^2-1)(x^2-y^2)$ and 
$Y_4^0 = c_4 \frac{1}{8} ( 35z^4 - 30z^2 + 3)$.
For state O, we note further that eq (\ref{L4O})
is symmetric under $x^2 \leftrightarrow y^2$,
thus $Y_4^2 + Y_4^{-2}$ is not involved
(as expected since its isotropy group contains the element
$C_4$).  Noting that $Y_4^0$ is cylindrically
symmetric but $Y_4^4 + Y_4^{-4}$ is not,
we find easily that the expression eq (\ref{L4O})
can be rewritten as
\begin{displaymath}
- \frac{1}{c_4} [ Y_4^0 + \sqrt{\frac{5}{14}} (Y_4^4+Y_4^{-4}) ]
\end{displaymath}
and hence the spin-wavefunction eq (\ref{PsiO4}).

For the state C,
it is more convenient to work with
\begin{equation}
\Psi' = 7 (\epsilon y^2z^2 + \epsilon^2 z^2 x^2 + x^2 y^2) 
  + (\epsilon x^2 + \epsilon^2 y^2 + z^2)  \ .
\end{equation}
Using the definition of $\epsilon$, we get,
\begin{equation}
\Psi' = 7 \left( - \frac{(x^2 +y^2)z^2}{2}
   - \frac{\sqrt{3}}{2} i z^2 (x^2-y^2) + x^2 y^2  \right)
  + \left( -\frac{x^2+y^2}{2} 
   + \frac{\sqrt{3}}{2} i (x^2-y^2) + z^2 \right)  \ .
\end{equation}
Our task is again simplified by noting that 
$Y_4^2+Y_4^{-2}$ is the only term that is asymmetric
under $ x \leftrightarrow y$, 
$Y_4^4+Y_4^{-4}$ is the only term that is symmetric
under $x \leftrightarrow y$ yet not cylindrically
symmetric.  We can then find easily
\begin{equation}
\Psi' = \frac{1}{c_4} [ Y_4^0 
  - \sqrt{\frac{6}{5}} i ( Y_4^2+Y_4^{-2}) -
  \sqrt{\frac{7}{10}} (Y_4^4 + Y_4^{-4} ) ]
\end{equation}
hence
the wavefunction is proportional to
$(-\sqrt{\frac{7}{10}}, 0, 
  -\sqrt{\frac{6}{5}}i, 0, 
 1,0, -\sqrt{\frac{6}{5}}i, 0, -\sqrt{\frac{7}{10}})$
and therefore the expression (\ref{PsiC4}) as shown in text.

\end{document}